\documentclass[10pt,showkeys,showpacs,amsmath,amssymb,prl]{revtex4}
\usepackage[english]{babel}
\usepackage[T2A]{fontenc}
\usepackage[cp866]{inputenc}
\usepackage[dvips]{graphicx}
\usepackage{amsmath,euscript,amssymb}
\newcommand{\be}{\begin{equation}}
\newcommand{\ee}{\end{equation}}
\newcommand{\bea}{\begin{eqnarray}}
\newcommand{\eea}{\end{eqnarray}}
\def\vh{\varphi}

\begin{document}

\title{Higgs effect in Conformal Cosmology \& Supernova Data}

\author{B.M. Barbashov}
\author{L.A. Glinka}\email{glinka@theor.jinr.ru}
\author{V.N. Pervushin}
\author{A.F. Zakharov}
\affiliation{Joint Institute for Nuclear Research, Dubna, Russia}

\date{\today}



\begin{abstract}

 The formulation of the Higgs effect is studied in
 the Glashow--Weinberg--Salam Standard Model, where
  the constant part of the Higgs potential is  identified with the zeroth
  mode of the Higgs field. In this model,
  the  Coleman--Weinberg effective potential obtained
   from the vacuum--vacuum
  transition amplitude is equal to unity at the extremum.
  This extremum immediately removes
 tremendous vacuum cosmological density and predicts mass of Higgs field. In this model,  the
 kinetic energy density of the Higgs field and any scalar field
 can be treated as the  rigid state origin that explains
 Supernova data  in the conformal
 cosmology without the $\Lambda$ term.
\end{abstract}

\maketitle

\section{Introduction}

The discovery and research of the Higgs boson are the highest
priority of modern elementary particle physics. Existing of the
Higgs particle is immediately associated with the fundamental
question about mass-generation mechanism \cite{hhg}. However, the
status of this mechanism in particle physics is not well and clear
defined. The Higgs potential is added to the Standard Model by
artificial phenomenological construction.
 This "by hand" construction leads to fundamental problems in correct formulation of
the theory - in this self-consistent spontaneous symmetry breaking
mechanism.

The other alive problem is the
 unification of the General Relativity with  the
 Standard Model leading to
 tremendous vacuum cosmological density. The problem of the vacuum
 density becomes
 crucial in the context of the explanation
 of the last supernovae data on the luminosity distance -- redshift
 relation \cite{Riess_04,Riess_98}.
 Recall that the best
fit to  the points including  189 high-redshift Type Ia
 supernovae~\protect \cite{Riess_04,Riess_98} requires a
cosmological constant $\Omega_{\Lambda}=0.7$ and $\Omega_{\rm Cold
Dark Matter}=0.3$ in the case of the standard cosmological model
with absolute units $ds^2_{\rm Einstein}=g_{\mu\nu}dx^\mu dx^\nu$.
An alternative
 description of the SN data without a $\Lambda$-term   are also discussed
 among other possibilities \cite{CC_papers,Behnke_02,Behnke_04,zakhy,CC-2}
as   evidence for the Weyl geometry \cite{we}, where all measurable
quantities and  their  units are considered on equal footing
$ds^2_{\rm Weyl}=ds^2_{\rm Einstein}/ds^2_{\rm Einstein~Units}$.
 The relative units leads to conformal
cosmological models, where
 the supernovae data
is compatible with the rigid state $\Omega_{rig}=0.85\pm 0.15$, and
$\Omega_{\rm CDM}=0.15\pm 0.10$. However, in both the cases, the
standard cosmology and the conformal one, a question arises ``What
is an origin of the $\Lambda$-term or the rigid state,
respectively?''

In this paper,  the  conformal-invariant formulation of the vacuum
Higgs effect is studied
 in the context of solution of the  problems of
 tremendous vacuum cosmological density and the origin of the
 rigid state needed for explanation of the supernovae data in
 conformal cosmology \cite{Behnke_02,Behnke_04}.
We show that the  origin of the rigid state can be kinetic energy
density of any scalar field  including the Higgs field, if the
potential energy of this field is suppressed or is absent.

As first we discuss the conformal-invariant formulation of the Higgs
effect and the condensate mechanism of spontaneously symmetry
breaking. Next part of paper is devoted to a status of the scalar
Higgs field in the context of the unification of the General
Relativity and the Standard Model compatible with the Newton law.
Finally we study the cosmological aspects of the Higgs effect.

\section{The Standard Model}
\subsection{Acceptable Higgs effect}

 The Standard Model constructed on  the Yang--Mills theory \cite{ym}
 with the symmetry group ${SU(2)}\times{U(1)}$ is known as the
Glashow-Weinberg-Salam theory of electroweak interactions
\cite{weak}.  The action of the Standard Model in the electroweak
sector with presence of the
 Higgs field can be written in the form
 \be\label{1-sm}
 S_{\rm SM}=\int d^4x
 {\cal L}_{\rm SM}=\int d^4x\left[{\cal L}_{\rm Ind}
 +{\cal L}_{\rm Higgs}
 \right],
 \ee
where
 \bea\label{M}
{\cal L}_{\rm Ind}=-\frac{1}{4}G^a_{\mu\nu}
G^{\mu\nu}_{a}-\frac{1}{4}F_{\mu\nu}
F^{\mu\nu}+\sum_s\bar{s}_1^{R}\imath \gamma^{\mu}\left(D_{\mu}^{(-)}
 +\imath
g^{\prime}B_{\mu}\right)s_1^{R}+\sum_s\bar{L}_s\imath
\gamma^{\mu}D^{(+)}_{\mu}L_s, \eea is the Higgs field independent
part of the Lagrangian and \bea\label{Ma}
{\mathcal{L}_{\rm{Higgs}}}= {\partial_\mu\phi\partial^\mu\phi}
-\phi\sum_sf_s\bar s s+\frac{{\phi^2}}{4}\sum_{\rm v} g^2_{\rm
v}V^2-
\underbrace{{\lambda}\left[\phi^2-\textrm{C}^2\right]^2}_{V_{\rm
Higgs}}
 \eea
is the Higgs field dependent
 part. Here
\begin{eqnarray}\label{1-6a}
 \sum_s f_s\bar s s&\equiv&\sum_{s=s_1,s_2} f_{s}\left[\bar s_{sR}s_{sL}
 +\bar s_{sL}s_{sR}\right],\\
\label{w1-6a} \frac{1}{4}\sum_{\rm v} g^2_{\rm
v}V^2&\equiv&\frac{1}{4}\sum_{\rm v=W_1,W_2,Z} g^2_{\rm
v}V^2=\frac{g^{2}}{4}W_\mu^{+}W^{-\mu}+\frac{g^{2}+g'^{2}}{4}Z_{\mu}Z^{\mu}
\end{eqnarray}
  are
  the mass-like terms  of  fermions  and W-,Z-bosons coupled with the Higgs field,
 \mbox{$G^a_{\mu\nu}=\partial_{\mu}A^{a}_{\nu}-\partial_{\nu}A^{a}_{\mu}+g\varepsilon_{abc}A^{b}_{\mu}A^{c}_{\nu}$}
is the field strength of non-Abelian $SU(2)$ fields (that give weak
interactions) and
\mbox{$F_{\mu\nu}=\partial_{\mu}B_{\nu}-\partial_{\nu} B_{\mu}$} is
the field strength of Abelian $U(1)$ (electromagnetic interaction)
ones,
$D_{\mu}^{(\pm)}=\partial_{\mu}-i{g}\frac{\tau_{a}}{2}A^a_{\mu}\pm\frac{i}{2}g^{\prime}B_{\mu}$
are the covariant derivatives, $\bar L_s=(\bar s_1^{L}\bar s_2^{L})$
are the fermion doublets,  $g$ and $g'$ are the Weinberg coupling
constants, and
 measurable gauge bosons $W^+_{\mu},~W^-_{\mu},~Z_{\mu}$ are defined by the relations:
\bea
W_{\mu}^{\pm}&\equiv&{A}_{\mu}^1\pm{A}_{\mu}^2={W}_{\mu}^1\pm{W}_{\mu}^2,\\
Z_{\mu}&\equiv&-B_{\mu}\sin\theta_{W}+A_{\mu}^3\cos\theta_{W},\\
\tan\theta_{W}&=&\frac{g'}{g},\eea where $\theta_{W}$ is the
Weinberg angle.

The crucial meaning has a distribution of the Higgs field $\phi$ on
the zeroth Fourier harmonic
 \be\label{h0-1}
 \langle\phi\rangle=\frac{1}{V_0}\int d^3x \phi
 \ee and the nonzeroth ones $h$, which we will call
 the Higgs boson
  \be\label{h-1}
 \phi=\langle\phi\rangle+\frac{h}{\sqrt{2}},~~~~\int d^3x h=0.
 \ee
In the acceptable  way, $\langle\phi\rangle$ satisfies the
 particle vacuum classical equation ($h=0$)
 \be\label{hc-1}
 \frac{\delta V_{\rm Higgs}(\langle\phi\rangle)}
 {\delta\langle\phi\rangle}=4{\langle\phi\rangle}
 [{\langle\phi\rangle}^2-\textrm{C}^2]=0
 \ee
 that has two solutions
\be\label{hc-2}
 {\langle\phi\rangle}_1=0,~~~~~~~~~~~~~~~
 {\langle\phi\rangle}_2=\textrm{C}\not =0.
 \ee
 The second solution corresponds to the spontaneous vacuum symmetry
 breaking
 that determines the masses of all elementary particles
 \bea
\label{0W-1}M_{W}&=&\frac{\langle\phi\rangle}{\sqrt{2}}g\\\label{0Z-1}
M_{Z}&=&\frac{\langle\phi\rangle}{\sqrt{2}}\sqrt{g^2+g'^2}
\\\label{0s-1}m_{s}&=& \langle\phi\rangle y_s,
\eea according to the definitions of the masses of vector (v) and
fermion (s) particles
 \be\label{vs-1}
{\cal L}_{\rm mass~ terms}= \frac{M_v^2}{2}V_\mu V^\mu-m_s\bar s s.
 \ee

\subsection{Conformal Higgs effect }

 The Hamiltonian approach to the Standard Model  considered in \cite{252} leads to
 fundamental operator quantization that allows
a possibility of  dynamic spontaneous symmetry breaking
 based on
the Higgs potential  (\ref{Ma}), where instead of a dimensional
parameter $\textrm{C}$ we substitute the zeroth Fourier harmonic
(\ref{h0-1}) \bea\label{0-Ma}{\mathcal{L}_{\rm{Higgs}}}&=&
{\partial_\mu\phi\partial^\mu\phi} -\phi\sum_sf_s\bar s
s+\frac{{\phi^2}}{4}\sum_{\rm v} g^2_{\rm v}V^2-
\underbrace{{\lambda}\left[\phi^2-\langle\phi
\rangle^2\right]^2}_{V_{\rm Higgs}}.
 \eea
 After the separation of the zeroth mode (\ref{h-1}) the
 bilinear part of the Higgs Lagrangian takes the form
 \bea\label{cMa}
 {\mathcal{L}_{\rm{Higgs}}^{\textrm{bilinear}}}&=& \frac{1}{2}{\partial_\mu h\partial^\mu h}
 -\langle\phi\rangle\sum_sf_s\bar s s+\frac{{\langle\phi\rangle^2}}{4}\sum_{\rm v} g^2_{\rm
 v}V^2- 2{\lambda}{\langle\phi\rangle^2}h^2. \eea
 In the lowest order in the coupling constant,
 the bilinear Lagrangian of the sum of all
 fields
\mbox{$S^{(2)}=\sum_F S_F^{(2)}[\langle\phi\rangle]$} arises with
the masses of vector (\ref{0W-1}), (\ref{0Z-1}), fermion (s)
(\ref{0s-1}) and
 Higgs (h) particles:
 \bea
\label{0h-1}
 m_h&=&2\sqrt{\lambda}\langle\,\phi\rangle.
\eea

 The sum of all
 vacuum-vacuum transition amplitude diagrams of the theory is known
 as the effective Coleman -- Weinberg
potential \cite{coleman79}
 \bea\label{eff-1}\textsf{V}^{\rm conf}_{eff}(\langle\,\phi\rangle)=-i
 \mathtt{Tr}\log <0|0>_{(\langle\,\phi\rangle)}=-i \mathtt{Tr}\log
  \prod_{F} G^{-A_F}_F[\langle\,\phi\rangle]G^{A_F}_F[\phi_{\rm I}],\eea
 where $G^{-A_F}_F$ are the Green-function operators with $A_F=1/2$ for bosons and
 $A_F=-1$ for fermions.
 In this case, the unit vacuum-vacuum transition amplitude $<0|0>
~\Big|_{\langle\,\phi\rangle =\phi_{\rm I}}=1$ means that
 \bea
 \label{va-2}
 \textsf{V}^{\rm conf}_{eff} (\phi_{\rm I})=0,
 \eea
where
   $\phi_{\rm
 I}$ is a  solution of the variation equation
\bea\label{va-1}
 &&\partial^2_0\langle\phi\rangle+\frac{d\textsf{V}^{\rm
 conf}_{eff}(\langle\,\phi\rangle)}{d\langle\,\phi\rangle}~
 \Bigg|_{\langle\,\phi\rangle=\phi_{\rm I}}
 =\partial^2_0\langle\phi\rangle+\sum_sf_s<\!\!\bar s
 s\!\!>-\frac{{\langle\,\phi\rangle}}{2}\sum_{\rm v} g^2_{\rm
 v}<\!\!V^2\!\!>\,+\,4\lambda{\langle\,\phi\rangle}<\!\!h^2\!\!>=0,
 \eea
 here $<V^2>,<\bar ss>,<h^2>$ are the condensates
 determined by   the Green
 functions in \cite{252}  \bea \label{v3-4-8}
 <V^2>=\langle V_\mu(x)V_\mu(x) \rangle &=&
 %-\frac{2}{(2\pi)^3}\int\frac{d^3\bf{k}}{\sqrt{M_{R\,\rm v}^2+\bf{k}^2}}=
 -2 L_{\rm v}^2(M_{R\,\rm v}),
 \\\label{s3-4-8}
 <\bar ss>=\langle \overline{\psi}_\alpha(x)  \psi_\alpha(x)\rangle&=&
 %\frac{m_s}{(2\pi)^3}\int\frac{d^3\bf{k}}{\sqrt{m_{R\,s}^2+\bf{k}^2}}=
 -m_{R\,s}L_s^2(m_{R\,s})
 ;\\
\label{h3-4-8}< h^2> =\langle h(x)h(x) \rangle &=&
 %-\frac{1}{2(2\pi)^3}\int\frac{d^3\bf{k}}{\sqrt{m^2_{R\,h}+\bf{k}^2}}=
 \frac{1}{2} L_h^2(m_{R\,h});
 \eea
here $L^2_{p}(m_p^2)$ are values of the integral \bea\label{v-5}
 L^2_{p}(m_p^2)=\frac{1}{(2\pi)^3}\int\frac{d^3\bf{k}}{\sqrt{m^2_p+\bf{k}^2}}.
 %\simeq\int \frac{d^3k}{|{\bf k}|}
% =\frac{k^2_{\rm max}}{(2\pi)^2}.%~~~~~(k_{\rm max}\simeq2\pi\vh_0).
 \eea
%that coincide in the large cut off momentum limit  \mbox{$L^2 \gg
%m^2_s,M^2_W$}

 Finally, using the definitions of the condensates and
 masses (\ref{0W-1}), (\ref{0Z-1}),(\ref{0s-1}),(\ref{0h-1}) we
 obtain the equation of motion
\bea\label{0-Ma-3}
\langle\phi\rangle\partial^2_0\langle\phi\rangle=\sum_sm_s^2
L_s^2-2\sum_{\rm v} M_{\rm v}^2L^2_{\rm v}-\frac{1}{2}m_h^2L^2_{\rm
h} . \eea
 In the class of  constant solutions $\partial^2_0\langle\phi\rangle\equiv 0$
 this equation has two solutions
\be\label{hc-2a}
 {\langle\phi\rangle}_1=0,~~~~~~~~~~~~~~~
 {\langle\phi\rangle}_2=\textrm{C}\not =0.
 \ee
 The nonzero solution means that there is
 the Gell-Mann--Oakes--Renner type relation
\bea\label{31-t}
 L_h^2 m^2_h=2\sum_{s=s_1,s_2}{L_s^2}m^2_s
 -4[2M_W^2L_W^2+M_Z^2L_Z^2].
 \eea
In the minimal SM \cite{db}, a three color t-quark dominates $\sum_s
m_s^2\simeq 3m_t^2$ because
  contributions of other fermions $\sum_{s\not=t} m_s^2/2m_t\sim 0.17$ GeV
 are very small. In the large cut-off limit we have the equality
 $L_W^2=L_Z^2=L_s^2=L_h^2=L^2$
 which immediately leads to calculation of the Higgs mass \bea\label{41-t}
 m_h=\sqrt{6m^2_t-4[2M_W^2+M_Z^2]}=311.6\pm 8.9 \mbox{\rm GeV},
 \eea
   if we substitute  the PDG data by the  values of masses of bosons
 $M_W=80.403\pm0.029$ GeV, $M_Z=91.1876\pm0.00021$ GeV,
 and t-quark $m_t=174.2\pm 3.3$ GeV.

 If we suppose that the condensates $L^2_p(m^2_{\rm R\,p})$
 are defined by the subtraction
 procedure associated with  the renormalization of masses and wave functions
 leading to the finite value
 \be\label{r-1}
 L^2_{\rm R p}(m^2_{\rm R\,p})=L^2_{\rm p}(m^2_{\rm Rp})-
 L^2_{\rm p}(0)-m^2_{\rm R p}
 \left[\frac{d}{d m^2_{\rm R\,p}}L^2_{\rm p}(m^2_{\rm Rp})
 \Bigg|_{m^2_{\rm R\,p}=0}\right]=
 \gamma m^2_{\rm R\,p},
 \ee
where $\gamma$ is the universal parameter,
 the sum rule (\ref{31-t}) takes the form
$%\bea\label{R-2}
  m^4_h=6m^4_t
 -4[2M_W^4+M_Z^4].
 $ %\eea
This sum rule gives the value of the Higgs mass
 \bea\label{R-3}
  m_h=\sqrt[4]{6m^4_t
 -4[2M_W^4+M_Z^4]}=264.6\pm 7.5 \mbox{\rm GeV}.
 \eea

\section{Conformal unification with the
General Relativity}

\subsection{The unified action}
 As it is known \cite{H} pure Einstein gravitation in absence
 of the matter fields is given by the Hilbert action
 \be\label{1-h1}
 S_{GR}=\int d^4x\sqrt{-g}
 \left[-\frac{\vh_0^2}{6}~^{(4)}R(g)\right],
 \ee
 where $1/\vh_0^2=8\pi G /3$
 is dimensional parameter determined by the Newton constant
$G$, and $g$ is the  metric determinant.

The
action of the SM in the electroweak sector, with presence of the
conformally coupled Higgs field can be write in the form
 \be\label{2-sm}
 S_{\rm SM}=\int d^4x\sqrt{-g}
 {\cal L}_{\rm SM}=\int d^4x\sqrt{-g}
 \left[\frac{\phi^2}{6}{}^{(4)}R(g)+{\cal L}_{\rm Ind}
 +{\cal L}_{\rm Higgs}
 \right],
 \ee
that differs from  (\ref{1-sm}) by the curvature term.

 In acceptable QFT
 unification of the General Relativity and the Standard Model
 is considered as the direct algebraical sum of GR (\ref{1-h1})
 and SM (\ref{1-sm}) actions \be\label{t-1}
 S_{\rm GR\&SM}=S_{\rm GR}+S_{\rm SM}.
 \ee
 in the Riemannian manifold.

\subsection{Higgs-metric mixing. Restoration of Newton's law.}

The General Relativity and the Standard Model  reflect almost all
physical effects and phenomena
 revealed by measurements and observations,
 however, it does not means that the direct sum
 of the actions of GR an SM lies in agreement with all these effects and
 phenomena.
 One can see that the conformal coupling Higgs field $\phi$
 with conformal weight $n=-1$
 distorts the Newton coupling constant in the Hilbert
  action (\ref{1-h1})
 \be\label{22-1}
 S_{\rm GR+Scalar}=\int d^4x\sqrt{-g}
 \left[-\left(1\!-\!\frac{\phi^2}{\vh_0^2}\right)\frac{\vh_0^2}{6}R(g)
 +g^{\mu\nu}\partial_\mu\phi\partial_\nu\phi
 \right]
 \ee
 due to
 the additional curvature term in the Higgs Lagrangian (\ref{Ma})
 $1\!-\!{\phi^2}/\vh_0^2$.
 This distortion
 changes
 the Einstein equations
 and their standard solutions
  of the Schwarzschild type and other \cite{B74,PPG,PS}.

   The coefficient $1\!-\!{\phi^2}/{\vh_0^2}$
 restricts  region, where the Higgs   field   is given,
  by the
 condition $\phi^2 < {\vh_0^2}$,
  because in other region $\phi^2 > {\vh_0^2}$
  the sign before the 4-dimensional curvature is changed in the
  Hilbert action  (\ref{1-h1}).

In order to keep the Einstein theory  (\ref{1-h1}),  one needs to
consider only the field configuration such that $\phi^2 <
{\vh_0^2}$.
  For this case one can introduce  new  variables by the Bekenstein--Wagoner transformation \cite{B74}
\bea\label{9-h11}
 g_{\mu\nu}&=&g_{\mu\nu}^{\rm (B)}\cosh^2Q\simeq g_{\mu\nu}^{\rm (B)},
  \\\label{9-h6}
 \phi^2&=&\vh_0^2\sinh^2Q\simeq\vh_0^2 Q^2, \\\label{s9-h6}
 s_{\rm (B)}&=&(\cosh{Q})^{-3/2} s
  \eea
 considered in \cite{PPG,PS}.
 These
 variables restore the initial Einstein--Hilbert action  (\ref{22-1})
 with the standard Newton law in the following way
 \be\label{22-2}
 S_{\rm GR+Scalar}=\vh_0^2\int d^4x\sqrt{-g_{\rm (B)}}
 \left[-\frac{R(g_{\rm (B)})}{6}+g_{\rm (B)}^{\mu\nu}
 \partial_\mu Q \partial_\nu Q\right].
 \ee
 Now it is clear that \emph{the  Bekenstein--Wagoner (BW)
 transformation converts the ''conformal  coupling'' Higgs field
 with the  weight $n=-1$ into
 the  ''minimal coupling''} scalar field $Q$ - an angle
 of the scalar -- scale mixing
 that looks like a {\it scalar graviton} with the conformal weight
 $n=0$.

 The Planck mass became one more  parameter
  of the Higgs Lagrangian,  so that
    the lowest order of the Lagrangian
    after  decomposition
 \be\label{hig-1}
 Q=\langle Q\rangle+\frac{h}{\vh_0\sqrt{2}}
 \ee
 over small constant $\langle Q\rangle\sim \textrm{C}/\vh_0\ll 1$
   reproduces  the acceptable Standard Model action  (\ref{0-Ma})
\begin{eqnarray}
\label{t-2small} {\mathcal{L}^{\rm \lambda}_{\rm{Higgs}}}(\langle
Q\rangle)=\vh_0^2g_{\rm (B)}^{\mu\nu}
 \partial_\mu Q \partial_\nu Q-\langle Q\rangle\vh_0\sum_s f_s \bar s_{\rm (B)} s_{\rm
(B)} +\frac{\langle Q\rangle^2\vh_0^2}{4}\sum_{\rm v} g^2_{\rm
v}V^2-4{\lambda}  \langle Q\rangle^2\vh^2_0h^2.
\end{eqnarray}

\section{Conformal Cosmology. The Higgs effect.}

Since the end of the XX century supernovae data has widespread
tested for all theoretical cosmological models. The main reason of
this is the fact that supernovae "standard candles" are still
unknown or absent \cite{Panagia_05}. Moreover, the first
observational conclusion about accelerating Universe and existence
of non-vanishing the $\Lambda$-term was done with the cosmological
SNe Ia data. Therefore, typically  standard (and alternative)
cosmological approaches are checked with the test.

Models of Conformal Cosmology are also discussed among other
possibilities \cite{CC_papers,Behnke_02,Behnke_04,CC-2}.
 Conformal Cosmology is an alternative description of the Supernovae data without
the $\Lambda$-term as evidence for  the Weyl geometry \cite{we} with
the relative units interval $ds^2_{\rm Weyl}=ds^2_{\rm
Einstein}/ds^2_{\rm Einstein~Units}$ where all measurable quantities
and their  units are considered on equal footing.
%$$
%ds^2_{\rm relative~units}=\frac{ds^2_{\rm GR}}{ds^2_{\rm units}}
%$$
%in its
 There is the scalar version of the Weyl geometry
  described by the conformal-invariant action of
a massless scalar field \cite{pct} with the negative sign
 that is mathematically equivalent to the Hilbert action of the  General
 Relativity where the role of the scalar field $\phi$ is played by the
  parameter of the scale
transformation $g^{\Omega}=\Omega^2 g$ multiplied by the Planck mass
$\phi=
 \Omega M_{\rm Planck}\sqrt{{3}/{(8\pi)}}$ \cite{kl}.
 %$$S_{\rm Hilbert-Einstein}[g^{\Omega}=\Omega^2 g]=
% -S_{\rm Penrose-Chernikov-Tagirov}\left[\phi=
% \Omega M_{\rm Planck}\sqrt{\frac{3}{8\pi}}\right]$$.

  Let us consider the Friedmann--Robertson--Walker (FRW) metric
 \bea\label{zm-2}
 ds^2&=& ds^2_{\rm FRW}=a^2(x^0)[(d\eta)^2 -(dx^idx^i)],~~ds^2_{\rm
 Weyl}=(d\eta)^2 -(dx^idx^i),
 \\\label{zm-3}
  a&=&\vh/\vh_0,~~~\vh_0=M_{\rm Planck}\sqrt{\frac{3}{(8\pi)}}\\\label{zm-4}
  \eta&=&\int dx^0 N_0(x^0), ~~V_0=\int d^3x.
 \eea
 The conformal vacuum Higgs effect  in the  cosmological
 approximation is described by the action
 \be \label{zm-1a}
  S_{\rm c}[\vh|\overline{Q}]\Big|_{N_0\not =1}=
 V_0\!\int\! dx^0  \!
 \left[ \frac{\vh^2}{{N}_0}\left(\frac{d \langle
 Q\rangle}{dx^0}\right)^2-\frac{1}{{N}_0}\left(\frac{d
 \vh}{dx^0}\right)^2-{N}_0\textsf{V}^{\rm
 conf}_{eff}(\vh \langle\,Q\rangle)\right],
 \ee
 where $\textsf{V}^{\rm conf}_{eff}(\vh \langle\,Q\rangle)$
 is the Coleman--Weinberg effective potential and is given by the
 formula (\ref{eff-1}). These action and interval keep the symmetry
  with respect to
  reparametrizations of the  coordinate evolution
 parameter $x^0~\to~\overline{x}^0=\overline{x}^0(x^0)$.
 Therefore, the cosmological model  (\ref{zm-1a}) can be considered
 by analogy with
 a model of a relativistic particle in the Special Relativity (SR)
 including the Hamiltonian approach to this theory.
 The canonical conjugate momenta of the theory (\ref{zm-1a}) are
 \be\label{pp-1} P_\vh=2\vh'V_0,
~~~P_{\langle{Q}\rangle}=2\vh^2\langle{Q}\rangle'V_0\ee where
$f'=\dfrac{df}{d\eta}$. After Hamiltonian reduction the action has a
form \be\label{ham}
 S[\vh|\overline{Q}]=\int
 dx^0\left[P_{\langle Q\rangle}\frac{d \langle
 Q\rangle}{dx^0}-P_\vh \frac{d\vh}{dx^0}-\frac{N_0}{4V_0}\left(-P^2_\vh+E^2_\varphi\right)\right],
 \ee
 where
 \be\label{ev1}
 E_\varphi=2V_0\left[\frac{P_{\langle{Q}\rangle}^2}{4V_0^2\vh^2}+\textsf{V}^{\rm
 conf}_{eff}(\vh \langle{Q}\rangle)\right]^{1/2}
 \ee
 is treated as the {\it"energy of a universe"}.

 The classical energy constraint in the model (\ref{ham}) is
 \bea\label{2.3-11}
 P_\vh^2-E_\vh^2=0
\eea and repeat completely the cosmological equations
  in the case of
 the rigid state equation $\Omega_{\rm rigid}=1$ because
 due to the unit vacuum-vacuum transition amplitude $V^{\rm
 conf}_{eff}(\vh \langle\,Q_I\rangle)=0$
\bea\label{2.3-12}
 \vh_0^2 a'^2=\frac{P_{Q_I}^2}{4V_0^2\vh^2}
 =H_0^2 \frac{\Omega_{\rm rigid}}{a^2},
 \eea
  where $P_{Q_I}$ is a constant of the motion
 \be\label{2.3-14a}
  P_{Q_I}'=0,
  \ee
 because in the equation of motion
\be \label{2.3-14} P\,'_{\langle{Q}\rangle}+\frac{d\textsf{V}^{\rm
 conf}_{eff}(\vh \langle{Q}\rangle)}{d\langle{Q}\rangle}=0, \ee
 the last term is equal zero
  $\dfrac{d\textsf{V}^{\rm
 conf}_{eff}(\vh \langle{Q}\rangle)}{d\langle{Q}\rangle}=0$
if all masses satisfy the Gell-Mann--Oakes--Renner type relation
(\ref{31-t}).

 The solution of these equations take the form
\bea\label{2.3-16}
 \vh(\eta)=\vh_I\sqrt{1+2{\cal H}_I\eta}, ~~~~
 \langle{Q}\rangle(\eta)=Q_I+\log {\sqrt{1+2{\cal H}_I\eta}},
\eea
 where
\bea\label{2.3-17}
 \vh_I&=&\vh(\eta=0),~~~~~{\cal
 H}_I\equiv\frac{\vh'(\eta=0)}{\vh(\eta=0)}=
 \frac{P_{\langle{Q}\rangle}}{2V_0\vh_I^2},\\\label{2.3-18}
  Q_I&=&{\langle{Q}\rangle}(\eta=0),~~~~P_{\langle{Q}\rangle}={\rm const}
\eea
 are the ordinary ``free'' initial data of the equation of
 the motion.
 Besides of the Higgs field $Q_H$ can be one more
  (massless) scalar field $Q_A$ (of the type of axion (A)).
 In this case
\bea\label{2.3-16a}
 \vh(\eta)&=&\vh_I\sqrt{1+2{\cal H}_I\eta}, \\
  Q_A(\eta)&=&Q_{AI}+\frac{{\cal H}_A}{2{\cal H}_I}\log {(1+2{\cal
  H}_I\eta)},\\
Q_H(\eta)&=&Q_{HI}+\frac{{\cal H}_H}{2{\cal H}_I}\log {(1+2{\cal
  H}_I\eta)},
\eea where $\vh_I,{\cal H}_I=\sqrt{{\cal H}_A^2+{\cal H}_H^2}$ and
$Q_{AI},{\cal H}_A=
 \dfrac{P_A}{2V_0\vh_I^2}$ and $Q_{HI},{\cal H}_H=
 \dfrac{P_H}{2V_0\vh_I^2}$ are free initial data in the CMB frame of
 reference, in the contrast to the Inflationary model, where
 $\vh_I={\cal H}(\eta=\eta_0)$.

 As it was shown \cite{CC_papers,gpk} there are initial data of
quantum
 creation of matter at
$z_I+1= 10^{15}/3$, and a value of the Higgs-metric mixing ``angle''
$Q_0\simeq3 \times 10^{-17}$
  is in agreement with the present-day energy budget of the Universe.

\section{Summary}

In this paper we study the  conformal-invariant formulation of the
vacuum Higgs effect
 in the context of solution of the  problems of
 tremendous vacuum cosmological density and the origin of the
 rigid state needed for explanation of the Supernova data in
 conformal cosmology \cite{Behnke_02,Behnke_04}.

 The conformal formulation of General
 Relativity  and Standard Model
  was studied in the context
 of its consistency with the Higgs effect,
 Newton law,  and
  the Super Novae luminosity-distance redshift
 relation.
The conformal formulation admits the dynamic version of the Higgs
potential, where its constant parameter is replaced by the zeroth
Fourier harmonic of the very Higgs field. In this case, the
  zeroth mode  equation is a new sum-rule that
 predicts mass of the Higgs field (\ref{41-t}) or (\ref{R-3}).

In this way  the conformal vacuum version of the Higgs mechanism
gives the possibility to solve the question about the nonzero vacuum
value of the scalar field with the zero vacuum cosmological energy
as a consequence of the unit vacuum-vacuum transition amplitude. The
conformal vacuum version of the Higgs mechanism gives the
possibility to explain the dominance of the most singular rigid
state at the  epoch of the intensive vacuum creation of the
primordial vector bosons. As it was shown, it is compatible with
energy budget of the universe \cite{CC_papers} and the Supernova
data \cite{Behnke_02,Behnke_04,zakhy}.

\section{Acknowledgements}

Authors are grateful to K.A. Bronnikov, P.H. Chankowski, V.V.
Kassandrov, D.I. Kazakov, E.A. Kuraev, Z. Lalak, D.G. Pavlov, S.
Pokorski, Yu.P. Rybakov and I. Tkachev for interesting and critical
discussions.

\end{document}